\documentclass[prb,showpacs,12pt]{revtex4}
\usepackage{epsfig}
\begin{document}

\title{Crystal field theory of Co$^{2+}$ in doped ZnO}

\author{R.O. Kuzian}
\affiliation{  Institute for Materials Science, Krzhizhanovskogo 3,
03180, Kiev, Ukraine}
\author{A.M. Dar{\'e}}
\affiliation{Laboratoire Mat{\'e}riaux et Micro{\'e}lectronique de
Provence, Facult\'e des Sciences et Techniques de St J\'er\^ome,
13397 Marseille Cedex 20, France}
\author{P. Sati}
\affiliation{Laboratoire Mat{\'e}riaux et Micro{\'e}lectronique de
Provence, Facult\'e des Sciences et Techniques de St J\'er\^ome,
13397 Marseille Cedex 20, France}
\author{R. Hayn}
\affiliation{Laboratoire Mat{\'e}riaux et Micro{\'e}lectronique de
Provence, Facult\'e des Sciences et Techniques de St J\'er\^ome,
13397 Marseille Cedex 20, France}

\date{\today}

\begin{abstract}
We present a crystal field theory of transition metal impurities in
semiconductors in a trigonally distorted tetrahedral coordination.
We develop a perturbative scheme to treat covalency effects within
the weak ligand field case (Coulomb interaction dominates over
one-particle splitting) and apply it to ZnO:Co$^{2+}$ (3$d^7$).
Using the large value of the charge transfer energy $\Delta _{pd}$
compared to the $p$-$d$ hoppings, we perform a canonical
transformation which eliminates the coupling with ligands to first
order. As a result, we obtain an effective single-ion Hamiltonian,
where the influence of the ligands is reduced to the one-particle
'crystal field' acting on $d$-like-functions. This derivation allows
to elucidate the microscopic origin of various 'crystal field'
parameters and covalency reduction factors which are usually used
empirically for the interpretation of optical and electron spin
resonance experiments. The connection of these parameters with the
geometry of the local environment becomes transparent. The
experimentally known $g$-values and the zero-field splitting $2D$
are very well reproduced by the exact diagonalization of the
effective single-ion Hamiltonian with only one adjustable parameter
$\Delta_{pd}$. Alternatively to the numerical diagonalization we use
perturbation theory in the weak field scheme (Coulomb interaction
$\gg$ cubic splitting $\gg$ trigonal splitting and spin-orbit
coupling) to derive compact analytical expressions for the
spin-Hamiltonian parameters that reproduce the result of exact
diagonalization within 20\% of accuracy.
\end{abstract}


\maketitle
\section{Introduction}
The A$^{II}$B$^{VI}$ and A$^{III}$B$^{V}$ compounds with transition
metal impurities are called diluted magnetic semiconductors (DMS).
These systems are interesting both from practical and fundamental
points of view. Recently, they have attracted great interest as
potential materials for spintronics, the current trend of
electronics which manipulates also the spin of the carrier, not only
its charge. The challenge to physicists is the highly correlated
subsystem of transition metal ions (TMI) demonstrating peculiar
interplay of the covalency and of the localized nature of
$d$-electrons.

Usually, the results of optical and electron spin resonance (ESR)
experiments are phenomenologically
described\cite{Weakliem,Koidl,abragam} in terms of crystal field
(CF)\cite{Bethe,Sugano,Griffith} theory. It considers the one-site
Hamiltonian for the $d$-shell of TMI, where Coulomb interaction and
spin-orbit (SO) coupling terms are added to one particle CF term of
unspecified nature. The CF interpretation reflects the symmetry of
TMI environment, but the values of energetic parameters are taken
from experiment, and, thus, say nothing about physical interactions
behind them. It concerns not only the one-particle CF terms:
phenomenological reduction factors for Racah's Coulomb parameters
$B$ and $C$, for the orbital angular momentum, and for the SO
coupling are always introduced.

Within the phenomenological CF theory, the hierarchy of interactions
is well established in most complexes containing the transition
metal ion surrounded by ligands. For commonly met octahedral and
tetrahedral environments, the CF terms are divided into a large
cubic part and much smaller low symmetry terms. For the 3$d$ ions
the SO term is one or two orders of magnitude smaller than the cubic
terms. To decide the order of the interaction strength between
Coulomb forces and cubic CF is much more difficult. One
distinguishes the 'strong' and 'weak' CF cases.\cite{intermed} In
the 'strong' field approach,\cite{Sugano} the cubic Hamiltonian is
diagonalized first, and its eigenfunctions serve as a basis for the
subsequent consideration of Coulomb and remaining terms. In the
'weak' field approach, the many-body eigenfunctions of the
single-ion Hamiltonian are exposed to the action of CF.

The Hartree-Fock molecular orbital theory enables to take into
account the covalency within the 'strong' field
approach.\cite{Sugano} Then, in principle, the calculation of CF
parameters becomes possible, the origin of the reduction factors
becomes clear and they can be estimated. We mention here the simple
analytic approach proposed in Ref. \onlinecite{kuzmin91} for the CF
parameters calculation. It is based on Harrison's parametrization of
the electronic structure of solids\cite{Harrison} and it enables to
relate the CF parameters with the structure of the TMI environment.

In the 'weak' field case, the account of covalency involves the
Heitler-London configuration interaction (CI) approach.\cite{Sugano}
It is much more complicated as it works with the enlarged Hilbert
space of many-body functions. On this way, the description of the
DMS on the energy scale of the cubic splitting ($\sim 0.5$ eV) was
achieved,\cite{Mizokawa} but more fine properties (such as magnetic
anisotropy) were not obtained, and the relation of CI approach with
the phenomenological CF theory was not established.

The A$^{II}$B$^{VI}$ semiconductors crystallize in zincblende or
wurtzite structures. In both structures, the TMI substituting for
the A site has similar tetrahedral coordination. The A site has
cubic point symmetry in zincblende structure, and trigonal symmetry
in wurtzite structure. In CF theory, the cubic field of the
zincblende structure is described by one parameter $\Delta $, and
the trigonal field (wurtzite structure) by three parameters $\Delta
$, $\upsilon $, $\upsilon ^{\prime }$. The deviation from cubic
symmetry in wurtzite structure, though small ($\Delta \gg \upsilon
,\upsilon ^{\prime }$), has crucial influence on the magnetic
properties of TMI. The Co$^{2+}$ ion in trigonally distorted
environment displays a large single-ion magnetic anisotropy, which
strongly affects the magnetic properties of Co-based
DMS.\cite{Sati05} The anisotropy is linear in the trigonal field
parameters $\upsilon$ and $\upsilon^{\prime}$. The empirical
determination of $\upsilon $ and $\upsilon ^{\prime }$ is difficult
and ambiguous. In optical transitions they are masked by a large
$\Delta $, and various sets give similar
description,\cite{Koidl,MacFarlane} their relation with the TMI
environment structure remains unclear. It is thus highly desirable
to have a microscopic model that would at least diminish the number
of adjustable parameters.

One microscopic model of CF is well known and was the first one that
appeared with the CF concept itself. It is the point charge model
(PCM).\cite{Bethe} In this model, it is possible to calculate CF
both in 'strong' and 'weak' field approaches, but the CF is then
strongly underestimated because the hybridization with ligands is
neglected. The authors of Ref. \onlinecite{mao} have tried to
improve the PCM and proposed to modify the TMI $d$-function in such
a way that it gives the correct value of $\Delta $ within the PCM. In
fact, they changed one set of parameters (CF matrix elements) to
another one (those describing their Slater-type $d$-function). In
order to explain the reduction factors, they additionally introduced
phenomenologically the hybridization with ligands that changes the
Coulomb and SO interactions but does not contribute to the CF
splitting.

Nevertheless, at present time, any modification of the PCM cannot be
accepted as a physical model of the TMI in DMS. It is not adequate
even for an isolated impurity, e.g. it cannot explain the increase
of cubic splitting with the increase of covalency. Nor, it cannot be
the starting point for studies of the interaction of TMI with the
host valence band ($p-d$ exchange) and between the impurities
(superexchange). Both phenomena are due to the virtual hoppings of
electrons between ligand and TMI described by the $p-d$
Hamiltonian.\cite{Mizokawa}

In this work we establish the relation between the configuration
interaction (CI) approach and the phenomenology of the CF. Below we
will show that the many-body perturbation theory gives the
possibility to consider the $d$-level splitting by covalency in the
'weak' field case at the same level of simplicity and physical
transparency as for the pure electrostatic case (PCM). Our final
goal is the derivation of the effective low energy spin-Hamiltonian,
i.e. an effective Hamiltonian that describes the ground state
manifold response to the applied magnetic field.  We will
demonstrate with an example for the $3d^7$ ion in tetrahedral
coordination that spin Hamiltonian may be derived from realistic
many-body Hamiltonian by means of a set of canonical
transformations.

In a first step, one eliminates the coupling with ligands: i.e. we
pass from realistic $p-d$ Hamiltonian to CF-like Hamiltonian for
$d$-ion only (Section II.A). We find a renormalization (i.e.\
reduction) of Coulomb and SO parameters (Section II.B).  We
generalize the ideas of Ref. \onlinecite{kuzmin91} and give the
expressions for CF parameters via $d-p$ hopping and charge transfer
energy difference. It is important to note that these expressions
give the connection of CF parameters with the structure of the TMI
environment (Section III).

Having derived the parameters for the effective CF-like Hamiltonian
in Section III, we calculate then in a second step the zero-field
splitting $2D$ and the $g$-factors, i.e.\ the parameters of the
low-energy spin-Hamiltonian. The basis for the crystal field
Hamiltonian of a 3$d^7$ configuration including SO coupling has 120
dimensions and will be diagonalized exactly (details are given in
the Appendix). It follows that we do not have to worry about the
'weak' field hypothesis for this effective Hamiltonian: indeed this
procedure is equally valid if cubic splitting $\Delta$ and Coulomb
interaction parameters $B$ and $C$ would be of the same order.

On the other hand, in the given case of ZnO:Co, we can alternatively
obtain an analytical closed expression which connects the parameters
of the microscopic Hamiltonian with the parameters of the effective
spin-Hamiltonian. For that purpose we use perturbation theory in the
spirit of the 'weak' field scheme. We construct the many-body basis
for a $d^7$ ion (Section IV.A) and make two subsequent
transformations (Section IV.B). The first one takes into account the
fact that the cubic splitting is smaller than the remaining Coulomb
interaction, and the second treats the trigonal coupling as a
perturbation. In the final step we take into account spin-orbit
coupling in order to obtain the spin-Hamiltonian observed in ESR.

In the Section V, we apply our approach to the case of ZnO:Co and we
compare the numerical diagonalization results with the perturbative
formula finding reasonable agreement. We compare also our parameter
set which was calculated microscopically with the phenomenological
parameter sets derived before from optics and ESR.

\section{Microscopic foundation of crystal field theory}

\subsection{Crystal field Hamiltonian}

In this section we show that starting from realistic $p-d$ Hamiltonian
it is possible to derive an effective single-ion Hamiltonian. The
latter has the form of a classical crystal field Hamiltonian. The
crystal field parameters acquire a clear microscopic meaning and the
essential property of being calculable and connected with the
structure of the TMI surroundings. The main point is the large energy
separation of configurations $d^n$ and $d^{n+1}L$ (respectively
$n$ electrons in $d$-shell of the TMI and
$n+1$ $d$-electrons plus one hole in the valence band),
compared with the TMI-ligand hopping.\cite{kuzmin91}

The appropriate Hamiltonian should be written in the basis of
spherically symmetric functions. The basis should not necessarily
coincide with the one for free ions. A more contracted basis is
suitable for solids, e.g.\ that one of the FPLO method.\cite{FPLO}
Without specifying it, we will assume the existence of an
orthonormal basis of one-particle spherically symmetric functions
localized on lattice sites. We assume also that it is a 'minimal'
basis set, i.e. one radial function suffices for the description of
one electronic shell. The existence of such a basis is not
explicitly proved, but there are indirect evidences in favor of it.
First, the FPLO method enables to explicitly construct
non-orthogonal minimal basis of localized functions. Second, W.A.
Harrison succeeded in describing the electronic structure of a huge
number of compounds, assuming the existence of such basis, and
empirically fitting the Hamiltonian matrix elements in this
basis.\cite{Harrison} The Hamiltonian may be written as
\begin{equation}  \label{Hpd}
\hat H=\hat H_d+\hat H_p+\hat T_{pd} \ ,
\end{equation}
where  $\hat H_d,\hat H_p$ are local Hamiltonians for the TMI and
ligands, respectively, $\hat T_{pd}$ describes electron hoppings
between the TMI and ligands. In the superposition model,
\cite{Newman} the contributions of separate ligands are superimposed
above each other. Further on, we will give a foundation for that
rule, following the lines given in Ref. \onlinecite{kuzmin91}. So,
it suffices to consider the ligand at the point $(0,0,R)$. Then, the
generalization to another geometry is straightforward.

In the zero-order Hamiltonian we include the diagonal one-particle
terms and dominant Coulomb interaction
\begin{equation}  \label{H0}
\hat H_0=\epsilon _d \hat N_d+\epsilon _p \hat N_p+\hat U_d+\hat
U_p \ ,
\end{equation}
where
$$
\hat N_l\equiv \sum_s \sum_{m=-l}^l\hat n_{m,s}\ ,\ \hat n_{m,s}=
c_{m,s}^{\dagger }c_{m,s}\ , \ \hat U_l=\frac{A_l}{2}\left(\hat
N_l^2-\hat N_l\right)  \  ,
$$
$\epsilon _d,\epsilon _p$ are the one-particle energies of $d$ and
$p$ states, $A_d$ and $A_p$ are the corresponding Racah's
parameters; the operator $c_{m,s}^{\dagger }=d_{m,s}^{\dagger
}(p_{m,s}^{\dagger })$ creates an electron with the one-particle
basis $d(p)$ wave function with angular momentum and spin
projections $m$ and $s$ on the TMI and on the ligand site
respectively. In the ground state of $H_0$, the $d$-shell of the TMI
contains $n$ electrons and the ligand has the closed $p$-shell with
$n_p=6$ electrons.
The hopping Hamiltonian
\begin{equation}  \label{Tpd}
\hat T_{pd}=\sum_{s}\sum_{m =-1 }^1 t_{pdm}\left( d_{m,s}^{\dagger
}p_{m,s}+p_{m,s}^{\dagger }d_{m,s}\right)
\end{equation}
couples configurations with different numbers of $d$-electrons. The
most important is the coupling between the configurations $d^np^6$
and  $d^{n+1}p^5$. The hybridization with the conduction band
depends on second nearest neighbor $d-s$ hopping matrix element and
may be neglected. In the two center approximation, the hopping $\hat
T_{pd}$ (Eq.(\ref{Tpd})) is diagonal over the angular momentum
projection indices $m$ and $m^{\prime }$ due to the symmetry with
respect to the TMI-ligand axis.
We perform a canonical transformation which eliminates the hopping to
first order
\begin{equation}  \label{trans}
\hat H_{eff}=\exp (-\hat W)\hat H\exp (\hat W)\approx \hat H +\left[
\hat H,\hat W\right]+ \frac 12 \left[\left[ \hat H,\hat
W\right],\hat W\right].
\end{equation}
We choose for this the operator $\hat W$ in the form
\begin{equation}  \label{W}
\hat W= -\frac{1}{\Delta _{pd}}\sum_{s}\sum_{m =-1 }^1 t_{pdm}\left(
d_{m,s}^{\dagger }p_{m,s}-p_{m,s}^{\dagger }d_{m,s}\right) \ ,
\end{equation}
where
\begin{equation}  \label{Deltapd}
\Delta _{pd}= nA_d-(n_p-1)A_p+\epsilon _d-\epsilon _p \ .
\end{equation}
With this choice of $\Delta _{pd}$, the coupling between the
$d^np^6$ and $d^{n+1}p^5$ configurations vanishes in the first order
operator $\left[ \hat H_0,\hat W\right]+\hat T_{pd}$. Neglecting the
coupling with the high-energy $d^{n+2}p^4$ configurations in the
second order operators
\begin{eqnarray}
\label{HWW} \frac 12 \left[\left[ \hat H_0,\hat W\right],\hat
W\right]&=& \frac 12 \sum_{s}\sum_{m =-1 }^1 \left\{
\left(\frac{t_{pdm}}{\Delta _{pd}}\right)^2 \left[ p_{m,s}^{\dagger
} \hat \Delta p_{m,s}-  d_{m,s}^{\dagger } \hat \Delta
d_{m,s}\right]\right. \\
\nonumber &+&\left.  d_{m,s}^{\dagger}
\left[\frac{t_{pdm}\left(A_d+A_p\right)} {\Delta _{pd}^2}\hat
T_{pd}\right] p_{m,s}+{\rm
h.c.}   \right\}, \\
\left[ \hat T_{pd},\hat W\right]&=& 2\sum_{s}\sum_{m =-1 }^1
\frac{t_{pdm}^2}{\Delta
_{pd}}\left(d_{m,s}^{\dagger }d_{m,s}-p_{m,s}^{\dagger }p_{m,s}\right),
\end{eqnarray}
where
$$\hat \Delta \equiv \epsilon _d-\epsilon _p+A_d\hat N_d-A_p\hat N_p \ ,$$
we end with the effective Hamiltonian for $d^np^6$ configuration
\begin{eqnarray}
\label{Heff} \hat H_{eff}&=&\hat H_0+\sum_{s}\sum_{m =-1 }^1
\frac{t_{pdm}^2}{\Delta
_{pd}}\left(d_{m,s}^{\dagger }d_{m,s}-p_{m,s}^{\dagger }p_{m,s}\right) \\
\nonumber
&=&
\epsilon _d \hat N_d+\hat U_d+\hat H_{CF}+{\rm const}\ ,
\end{eqnarray}
with
\begin{equation}
\label{HCF} \hat H_{CF} = \sum_{s}\sum_{m =-1 }^1
\frac{t_{pdm}^2}{\Delta _{pd}}d_{m,s}^{\dagger }d_{m,s} \ .
\end{equation}
In the last equality of Eq.\ (\ref{Heff}) we have taken advantage of
the fact that every state of our subspace includes the
non-degenerate closed $p$-shell of ligand for which $\langle
p^6|p_{m,s}^{\dagger }p_{m^{\prime },s}|p^6\rangle =\delta
_{m,m^{\prime }}$ and all terms concerning the ligand become
constant.
We have obtained an effective single-ion Hamiltonian (Eq.(\ref{Heff}))
where the action of ligand has been reduced to the one-particle
'crystal field' term $\hat H_{CF}$ (\ref{HCF}). Let us recall that
we have assumed that
\begin{equation}\label{tllDelta}
t_{pdm}\ll \Delta _{pd}\ .
\end{equation}
Within this assumption we may neglect the terms of third and higher
orders in $\hat H_{eff}$ in Eq. (\ref{trans}). It is easy to see
that the second order contributions from separate ligands simply sum
up when we do subsequent transformations to remove hopping to first
order between the TMI and the different ligands of the nearest
surroundings. Thus, the assumption (\ref{tllDelta}) gives the range
of validity for the superposition model in our case. Our approach
also implies that besides Eq. (\ref{tllDelta}), the characteristic
energies of the terms in $\hat H_d$ not included in $H_0$ in Eq.
(\ref{H0}) are also smaller than $\Delta _{pd}$. At first this
concerns the Coulomb energies
\begin{equation}\label{BCllDelta}
    15B \sim 3C \ll \Delta _{pd}\ .
\end{equation}
The main point here is that in addition to the one-particle energy
difference, $\Delta _{pd}$ contains also the largest Coulomb
parameter $A_d$  (see Eq.\ (\ref{Deltapd2HF}) below).

For the chosen axially symmetric geometry, the 'crystal field' is
diagonal with respect to the angular momentum projection $m$. For the
general relative positions of TMI and ligands the $\hat H_{CF}$ will
have the form\cite{kuzmin91}
\begin{eqnarray}
\label{HCFgen} \hat H_{CF}&=&\sum_{s}\sum_{m,m^{\prime } \ =-2
}^2V_{m m^{\prime }
}\ d_{m,s}^{\dagger }d_{m^{\prime },s} \\
  \label{Vmm}
V_{mm^{\prime }}&=&\sum_i\left\{ b_4(R_i)A_{mm^{\prime
}}Y_4^{m-m^{\prime }}\left( \theta _i,\phi _i\right)
+b_2(R_i)B_{mm^{\prime }}Y_2^{m-m^{\prime }}\left( \theta _i,\phi
_i\right) +b_0(R_i)\delta _{mm^{\prime }}\right\},
\end{eqnarray}
where
\[
A_{mm^{\prime }}=\left( -1\right) ^{m^{\prime }}\frac{5\sqrt{4\pi
}}{27} C_{-m^{\prime }m}^{224}C_{00}^{224},\ B_{mm^{\prime }}=\left(
-1\right) ^{m^{\prime }}\frac{\sqrt{4\pi }}5C_{-m^{\prime
}m}^{222}C_{00}^{222},
\]
the $C_{m_1m_2}^{j_1j_2J}=\langle j_1j_2m_1m_2|JM=m_1+m_2\rangle$
are the Clebsh-Gordan coefficients (see e.g. Refs.
\onlinecite{Messiah,abragam}), and the $Y_l^{m}$ are the spherical
harmonics. The coefficients
\begin{equation}
b_k(R_i)=\frac{2k+1}{5}\left\{ \frac{t_{pd\sigma }^{2}}{\Delta
_{pd}} +\left[ 2-\frac{k(k+1)}{6} \right] \frac{t_{pd\pi
}^{2}}{\Delta _{pd}}\right\} , \label{bk}
\end{equation}
depend only on the nature of ligand and TMI and on the distance
between them; the standard notations $m=\sigma ,\pi $ in Eq.
(\ref{bk}) correspond to $m=0, \pm 1$ respectively. The summation in
Eq. (\ref{Vmm}) goes over the ligand spherical coordinates
$R_i,\theta _i,\phi _i$. In the spirit of the superposition
model,\cite{Newman} the physical and geometrical informations are
separated in Eq. (\ref{Vmm}). Note that in the general case the
summation in Eq. (\ref{HCFgen}) runs over all $d$-states.

The idea to use Harrison's parametrization for the calculation of
hybridization contribution to CF and to Eq.\ (\ref{bk}) was
first proposed in Ref.\ \onlinecite{kuzmin91} from perturbative
approximate diagonalization of the {\em mean field one-particle
part} of the $p-d$ Hamiltonian. In this scheme the $\Delta _{pd}$
has the meaning of Hartree-Fock energies difference
\begin{eqnarray}
\label{DeltaHF}
\Delta _{pd, HF}=\varepsilon _{d,HF}-\varepsilon _{p,HF} \\
\label{eHF} \varepsilon _{l,HF}=\epsilon _l+A_l\left(n_l-1\right).
\end{eqnarray}
In this sense the approach of Ref. \onlinecite{kuzmin91} is close to
the 'strong' CF scheme.

In the spirit of 'strong' CF scheme, another approach was developed
in Ref. \onlinecite{Fazzio84}. There, the Coulomb electron-electron
interaction is taken into account {\em after} the diagonalization of
the one-particle mean-field Hamiltonian. It is thus rewritten in
terms of eigenfunctions of cubic CF. The applicability of Racah's
parametrization of Coulomb integrals is then questioned. In DMS, the
'strong' CF scheme fails. It is unable to explain the position of
incomplete $d$-shell below the Fermi level, because the mean-field
neglects the configuration interaction. The 'weak' CF scheme is free
from such difficulties and our considerations show the way to
account for covalency in this scheme.  Concluding this remark, let
us mention that comparing Eqs.\ (\ref{Deltapd}), (\ref{DeltaHF}) and
(\ref{eHF}) we see that
\begin{equation}\label{Deltapd2HF}
\Delta _{pd}=\Delta _{pd,HF}+A_d\ .
\end{equation}
This relation explains why the TMI $d$-level having the Hartree-Fock
mean-field energy lower than the ligand $p$-level (e.g. $\varepsilon
_{d,HF}=-17.77\ {\rm eV}$ for Co is lower than $\varepsilon
_{p,HF}=-16.77\ {\rm eV}$ for oxygen\cite{Harrison}) remains
incompletely filled. As we mentioned in the Introduction, the
difference between 'strong' and 'weak' CF schemes reflects the
difference of Hartree-Fock and Heitler-London ways of accounting for
the covalency, our consideration being close to the Heitler-London
approach.

\subsection{Renormalization of Coulomb, spin-orbit and Zeeman terms}

In the phenomenological CF theory, the covalency is accounted for by
introduction of reduction factors for Racah's parameters $B,C$ and
for orbital angular momentum matrix elements. The orbital moment
appears in the spin-orbit interaction $\hat H_{SO}$ and in the
Zeeman term
\begin{equation}\label{HZ}
\hat H_Z=\mu_B\left( g_s\hat{\bf S} + \hat{\bf L}\right){\bf B} \; ,
\end{equation}
where $g_{s}=2.0023$ is the Land\'e's factor and $\mu_B$ the Bohr's
magneton.

The canonical transformation (Eq. (\ref{trans})) changes the
many-body basis of the problem. For the sake of consistency we
should transform every additional term of the Hamiltonian as well as
any observable. The covalency reduction factors naturally occur as a
result of the canonical transformation of the corresponding
operators. The total spin operator $\hat{\bf S}$ commutes with the
canonical transformation operator $\hat W$ (\ref{W}) and remains
unchanged. Let us now demonstrate the appearance of an orbital
reduction factor $k$ ($\hat L \rightarrow k \hat L$, $k<1$) for the
spin-orbit term. We begin with the canonical transformation of
annihilation operators and with the ligand situated at $(0,0,R)$
\begin{eqnarray}
\nonumber \tilde d_{m,s}&\approx &d_{m,s} +\left[ d_{m,s},\hat
W\right]+ \frac 12 \left[\left[ d_{m,s},\hat W\right],\hat W\right]
\\
\label{transd}
 &=& \left( 1-\frac 12 \lambda _m^2\right)d_{m,s}-\lambda _m p_{m,s}\ , \\
\nonumber \tilde p_{m,s}&\approx &\left( 1-\frac 12 \lambda
_m^2\right)p_{m,s}+\lambda _m d_{m,s} \ ,
\end{eqnarray}
where $\lambda _m =t_{pdm}/\Delta _{pd}$. The apparent similarity of
Eq.\ (\ref{transd}) with molecular-orbital expression should not
mislead the reader. We recall that it works only in the subspace of
$d^np^6$ and $d^{n+1}p^5$ configurations and $\Delta _{pd}$
(\ref{Deltapd}) depends on the number of $d$ and $p$ electrons $n,\
n_p$.

Substituting the transformed annihilation and creation operators
into the second quantization expression for an
operator, we immediately obtain its transformed version. The spin-orbit
operator acquires the form
\begin{eqnarray}
\nonumber \hat H_{SO}&=&\frac{\xi _{d,0}}{2}\sum_{m,m\prime
,s,s\prime }\tilde d_{m,s}^{\dag }\  \mbox{\boldmath$L$}^{d
}_{m,m\prime } \ \mbox{\boldmath$ \sigma$}_{s,s\prime }\ \tilde
d_{m\prime ,s\prime }+\frac{\xi_{p,0}} {2}\sum_{m,m\prime ,s,s\prime
}\tilde p_{m,s}^{\dag } \ \mbox{\boldmath$ L$}^{p }_{m,m\prime }\
\mbox{\boldmath$\sigma$}_{s,s\prime }\  \tilde
p_{m\prime ,s\prime } \\
\label{SO} &\approx & \frac{\xi _{d,0}}{2}\sum_{m,m\prime ,s,s\prime
}\left( 1-\frac 12 \lambda _m^2\right)d_{m,s}^{\dag }\
\mbox{\boldmath$ L$}^{d }_{m,m\prime } \
\mbox{\boldmath$\sigma$}_{s,s\prime } \left( 1-\frac 12 \lambda
_{m\prime }^2\right)
d_{m\prime ,s\prime } \\
\nonumber &+& \frac{\xi _{p,0}}{2}\sum_{s,s\prime }\
\sum_{m,m\prime \ =-1 }^1\lambda _m  \lambda _{m\prime }\
d_{m,s}^{\dag }\ \mbox{\boldmath$ L$}^{p }_{m,m\prime }\
\mbox{\boldmath$\sigma$}_{s,s\prime }\ d_{m\prime ,s\prime }\ ,
\end{eqnarray}
where $\mbox{\boldmath$L$}^{p\ (d)}_{m,m\prime }$ is the angular
momentum matrix vector for a $L=2 (d)$ or $L=1 (p)$ particle;
$\mbox{\boldmath$\sigma$}$ is the Pauli matrix vector. The
spin-orbit coupling value $\xi _{d(p),0}$ is the free-ion (ligand)
one (see below). For the light ligands (e.g. oxygen) $\xi _{p,0}\ll
\xi _{d,0}$, and this term is usually neglected.

For the ligand situated in the point with spherical coordinates
$(\varphi ,\theta ,R)$ in crystallographic coordinates, we should
first go to the local coordinate system $X^{\prime }Y^{\prime
}Z^{\prime }$ with the $Z^{\prime }$ axis pointing towards the ligand.
The rotation is described by three Euler angles $\{\alpha =\varphi
,\beta =\theta , \gamma =0\}$ and the operators in the two systems are
related by the linear transformation
\begin{equation}\label{d2dpr}
d_{m,s}^{\prime }=\sum _{m_1} D_{m_1,m}(\alpha ,\beta , \gamma )
d_{m_1,s}\ .
\end{equation}
The matrices $D_{m_1,m}(\alpha ,\beta , \gamma
)=R^{(2)}_{m_1,m}(\alpha ,\beta , \gamma )$ describe the
transformation of spherical harmonics between two coordinate
systems. \cite{Messiah} After a canonical transformation in the
local coordinate system according to Eq.\ (\ref{transd}), we perform
a backward rotation and obtain
\begin{eqnarray}
\nonumber \tilde d_{m,s}&=&\sum _{m_1} D_{m_1
,m}(0,-\theta ,-\varphi ) \tilde d_{m_1,s}^{\prime } \\
\label{transdgen} &=& d_{m,s} +\sum _{m_1} D_{m_1,m}(0,-\theta
,-\varphi ) \left[-\lambda _{m_1}p_{m_1 ,s}-\frac{\lambda
_{m_1}^2}{2}\sum _{m_2} D_{m_2,m_1}(\varphi,\theta , 0) d_{m_2
,s}\right].
\end{eqnarray}
The summation over all ligands will give the final expression for
$\tilde d_{m,s}$. Generally it is very complicated, but for highly
symmetric surroundings, it recovers a form similar to Eq.
(\ref{transd}). For example, for the tetrahedrally coordinated TMI
we have
\begin{eqnarray} \label{transcub}
\tilde d_{t_{2g},s} &=& \left( 1-\frac 23 \lambda _{\sigma }^2-
\frac 49 \lambda _{\pi }^2\right)d_{t_{2g},s} -\sqrt{\frac 43
\lambda _{\sigma }^2-
\frac 89 \lambda _{\pi }^2}\ p_{t_{2g},s}\ ,\\
\nonumber \tilde d_{e_g,s} &=& \left( 1- \frac 43 \lambda _{\pi
}^2\right)d_{e_g,s} - \frac{2\sqrt{6}}{3} \lambda _{\pi }\
p_{e_g,s}\ ,
\end{eqnarray}
where the operator $d_{\Gamma ,s}$ annihilates the electron in the
state which transforms according to the irreducible representation
$\Gamma =e_g, t_{2g} $; $m=\sigma ,\pi $ again means $m=0, \pm 1$.
Instead of operators
$p_{m,s}$, an admixture of symmetric combinations of ligand
orbitals enters the Eqs. (\ref{transcub}).

In the ground state of tetrahedrally coordinated $d^7$ ion, the
$t_{2g}$ states are filled (by holes) and only the non-diagonal
matrix element of angular momentum $\langle t_{2g}\left|\hat
L\right|e_g\rangle$ enters the expression of the spin-Hamiltonian
parameters. It means that we may substitute
\begin{eqnarray}\label{redL}
\hat L &\rightarrow k &\hat L\ ,\ \\
\label{kred}
k &=& \left( 1-\frac 23 \lambda _{\sigma }^2- \frac 49
\lambda _{\pi }^2\right)\left( 1- \frac 43 \lambda _{\pi
}^2\right)\approx \left( 1-\frac 23 \lambda _{\sigma }^2-
\frac{16}{9} \lambda _{\pi }^2\right)\ ,
\end{eqnarray}
in the Zeeman term (\ref{HZ}) and
\begin{equation}\label{redxi}
\xi _{d,0} \rightarrow \xi _d = k \xi _{d,0}\ ,
\end{equation}
in the spin-orbit term.

The matrix elements of the Coulomb interaction that were not
included in $H_0$ (\ref{H0}), acquire prefactors that are product of
four terms $\left( 1-\frac 12 \lambda _m^2\right)$. Therefore, the
reduction factors for the Coulomb interaction may differ from those
of SO and angular momentum terms. Nevertheless, an approximate
estimate may be done by multiplying the free-ion Racah's parameters
$B_0$ and $C_0$ with $k^2$ from Eqn.\ (\ref{kred}), i.e.
\begin{equation}
B_0 \rightarrow B=k^2 B_0 \, , \qquad C_0 \rightarrow C=k^2 C_0 \; .
\end{equation}
Such an approach is completely sufficient for the given case of
ZnO:Co since the lowest $^4F$ multiplet is well separated from the
higher multiplets and, correspondingly, the parameters of the
effective spin-Hamiltonian ($g$-factors and zero-field splitting)
are not very sensitive to $B$ and $C$ (see Section V below).

\section{Crystal field theory for the tetrahedrally coordinated $d^7$ ion}

\begin{figure}[h]
    \begin{center}
      \includegraphics[bb=170 300 415 550 width=.3,height=.3\textheight]{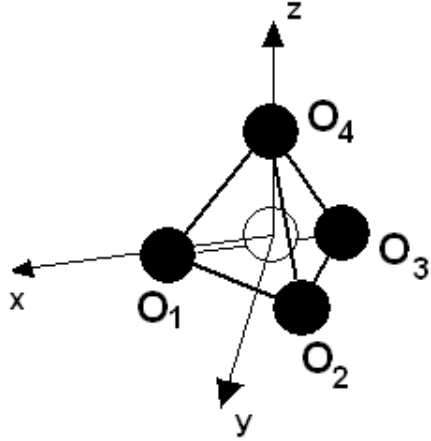}
      \caption{CoO$_4$ tetrahedron}\label{CoOtetrahedron}
\end{center}
\end{figure}

The effective single-ion Hamiltonian, derived in the previous
sections reads
\begin{equation}  \label{singleion}
\hat{H}_{eff}=\hat{H}_{Coul}+\hat{H}_{CF}+\hat{H}_{SO}+\hat{H}_{Z}\ .
\end{equation}
The Coulomb interaction within the $d$-shell $\hat{H}_{Coul}$ will
not be written down explicitly since it is diagonal in the many-body
basis corresponding to the weak field scheme.
In our approach, the CF parameters are connected with the local
environment of the TMI (see Table \ref{Tab1} and Figure
\ref{CoOtetrahedron} for the ZnO:Co example).
\begin{table}
\begin{tabular}{|c|c|c|c|}
\hline & $x/a$ & $y/a$ & $z/c$ \\ \hline
Co & 0 & 0 & 0  \\
O$_{1}$ & $\frac{1}{\sqrt{3}}$ & 0 & $-\frac{1}{8}+\delta $ \\
O$_{2}$ & $-\frac{1}{2\sqrt{3}}$ & $\frac{1}{2}$ & $-\frac{1}{8}+\delta $ \\
O$_{3}$ & $-\frac{1}{2\sqrt{3}}$ & $-\frac{1}{2}$ & $-\frac{1}{8}+\delta $ \\
O$_{4}$ & 0 & 0 & $\frac{3}{8}+\delta $ \\
\hline
\end{tabular}
\caption{Cartesian coordinates of the CoO$_{4}$ tetrahedron using
the lattice parameters $a=3.2427$~\AA , $c=5.1948$~\AA \ and $\delta
=0.0076$ of the host lattice ZnO.
\cite{c} \\[0pt] }\label{Tab1}
\end{table}
It is convenient to choose the zero of energy from the condition
Tr$V_{m,m\prime } = 5\sum_i b_0(R_i)=0$. The term $\hat{H}_{CF}$
in Eq. (\ref{HCFgen}) may be split into cubic and trigonal parts:
\begin{equation}  \label{eq2}
\hat{H}_{CF}=\hat{H}_{cub}+\hat{H}_{trig}\ ,
\end{equation}
since the ideal tetrahedron with $\frac{c}{a}=\sqrt{\frac{8}{3}}$
and $\delta =0$ is identical to cubic symmetry. In reality, however,
$\frac{c}{a}$ deviates from the ideal value and $\delta \neq 0$. The
trigonal field is described by three parameters. There exist several
different systems of notation in the literature and we will use here
the parameters $\Delta$, $\upsilon$, and $\upsilon ^{\prime}$ like
Koidl \cite{Koidl} and MacFarlane \cite{MacFarlane} (see also Ref.\
\onlinecite{Bates}). They are defined as a parametrization of the
crystal field matrix elements $V_{m m^\prime}$ (Eq. (\ref{HCFgen})) in the
one-particle basis of the trigonal coordinate system (the $z$ axis
is the threefold axis pointing towards O$_4$):
\begin{equation}\label{eq7}
\begin{array}{ll}
|x\rangle=\sqrt{\frac{2}{3}}|x^{2}-y^{2}\rangle-\sqrt{\frac{1}{3}}|zx\rangle\\
|y\rangle=-\sqrt{\frac{2}{3}}|xy\rangle-\sqrt{\frac{1}{3}}|zy\rangle\\
|z\rangle=|z^{2}\rangle\\
|v\rangle=\sqrt{\frac{1}{3}}|x^{2}-y^{2}\rangle+\sqrt{\frac{2}{3}}|zx\rangle\\
|w\rangle=-\sqrt{\frac{1}{3}}|xy\rangle+\sqrt{\frac{2}{3}}|zy\rangle\ ,\\
\end{array}
\end{equation}
where we have the usual real $d$-basis functions constructed out of
the complex  basis functions $|l\rangle=d^{+}_{l\sigma}|0\rangle$ on
the right hand side. The three basis functions $|x\rangle$,
$|y\rangle$ and $|z\rangle$ build up the $t_{2g}$ representation of
the cubic group and $|v\rangle$, $|w\rangle$ span up the $e_g$
subspace. The one-particle crystal field matrix elements
are given in this basis by:
\begin{equation}\label{eq31}
\begin{array}{ll}
V_{zz}=\frac{2}{5}\Delta-\frac{2\upsilon}{3}\\
V_{xx}=V_{yy}=\frac{2}{5}\Delta+\frac{\upsilon}{3}\\
V_{vv}=V_{ww}=-\frac{3}{5}\Delta\\
V_{xv}=\upsilon' \; .
\end{array}
\end{equation}
The relationship between the $\Delta$, $\upsilon$, and $\upsilon
^{\prime}$ parameters with the Stevens equivalent operators and
other parametrizations used in the literature can be found in the
Appendix.

If we substitute the ligand coordinates from Table \ref{Tab1} into $V_{m
m\prime }$ in Eq. (\ref{Vmm}) and then transform them into the trigonal
crystal field parameters we obtain
\begin{eqnarray}
\nonumber
\Delta &=& -\frac{5}{27}\left\{ b_4(R_1) \frac{72(z_1^2a^2-z_1^4)-3a^4+20
\sqrt{6}z_1a^3}{24R_1^4}-b_4(R_4)\right\}, \\
\nonumber
  \upsilon  &=& \frac{b_4(R_1)}{9R_1^4}\left\{ \frac{20}{7}
  \left[ 3(z_1^2a^2-z_1^4)-\frac{a^4}{8}\right] -
  \frac{5\sqrt{6}z_1a^3}{6} \right\}-\frac{20}{63}b_4(R_4) \\
  \label{xyz2Dvvp}
  & & - \frac{3}{14}\left[b_2(R_1)\frac{6z_1^2-a^2}{R_1^2}+2b_2(R_4)\right],\\
\nonumber
  \upsilon \prime  &=&\frac{5 b_4(R_1)}{9R_1^4}\left\{ \frac{\sqrt{2}}{7}
  \left[ 3(z_1^2a^2-z_1^4)-\frac{a^4}{8}\right]
  -\frac{\sqrt{3}z_1a^3}{12} \right\}-\frac{\sqrt{2}}{9}b_4(R_4) \\
\nonumber
  & & +
  \frac{\sqrt{2}}{14}\left[b_2(R_1)\frac{6z_1^2-a^2}{R_1^2}+2b_2(R_4)\right],
\end{eqnarray}
where $z_i$ is the $z$ coordinate of the ligand O$_i$;
$R_{1}=\sqrt{a^2/3+z_1^2}, \  R_4 = z_4$ are the corresponding
distances.

The importance of the non-diagonal matrix element between $e_g$ and
$t_{2g}$ states, $\upsilon^{\prime}$, was first pointed out in Refs.
\onlinecite{MacFarlane} and \onlinecite{MacFarlane67} and will be
outlined in the following, but it was not thoroughly treated in the
standard text books. \cite{abragam}
Hopping integrals $t_{pdm}$ have been calculated from Harrison's
table. \cite{Harrison} Then all the CF parameters depend only on one
value: $\Delta _{pd}$. Its determination is complicated by the fact
that the Coulomb repulsion $A_d$ is partially screened in
semiconductors and it differs much more from free ion value than $B$
and $C$. That is the reason why we have used $\Delta _{pd}$ as an
adjustable parameter. After determination of its value from
experimental knowledge of cubic parameter $\Delta $, the other
parameters $\upsilon$  and $\upsilon ^{\prime}$, are determined by
geometry.

For the spin-orbit and Zeeman terms we have
\begin{eqnarray}
\label{SO1ion}
  \hat H_{SO}&=&\frac{\xi _d}{2}\sum_{m,m\prime
,s,s\prime } d_{m,s}^{\dag } \mbox{\boldmath$L$}_{m,m\prime } \
\mbox{\boldmath$ \sigma$}_{s,s\prime }\
 d_{m\prime ,s\prime } \\
  \label{HZ1ion}
\hat H_Z &=& \mu_B \left( g_s\hat{\bf S} + k\hat{\bf L} \right){\bf B}\ ,
\end{eqnarray}
here $\xi _d$ is the renormalized spin-orbit coupling, and $k$ is
approximately given by Eq.\ (\ref{kred}); $\hat{\bf S}$ and
$\hat{\bf L}$ are respectively the total spin and orbital angular
momentum operators. Within the $^4F$ term, $H_{SO}$ may be rewritten
as
\begin{equation}\label{SO4F}
    \hat H_{SO,^4F}=\lambda \hat{\bf S}\hat{\bf L}\ ,\ \ {\rm with} \ \  \lambda =-\xi _d/3\ .
\end{equation}

\section{Spin Hamiltonian for the ground state manifold}

As it was already noted in the Introduction, this second step of our
calculation can be performed exactly by numerical diagonalization of
the effective single-ion Hamiltonian (\ref{singleion}) once we have
determined all its parameters (see Appendix). In that sense it is
not restricted to the 'weak' field case. In addition to the
numerical diagonalization, we derive below an analytic perturbative
formula which is valid in the 'weak' field case of ZnO:Co.

\subsection{Many-body basis}

\begin{figure}[h]
    \begin{center}
      \includegraphics[bb=70 140 515 735 width=.65,height=.65\textheight]{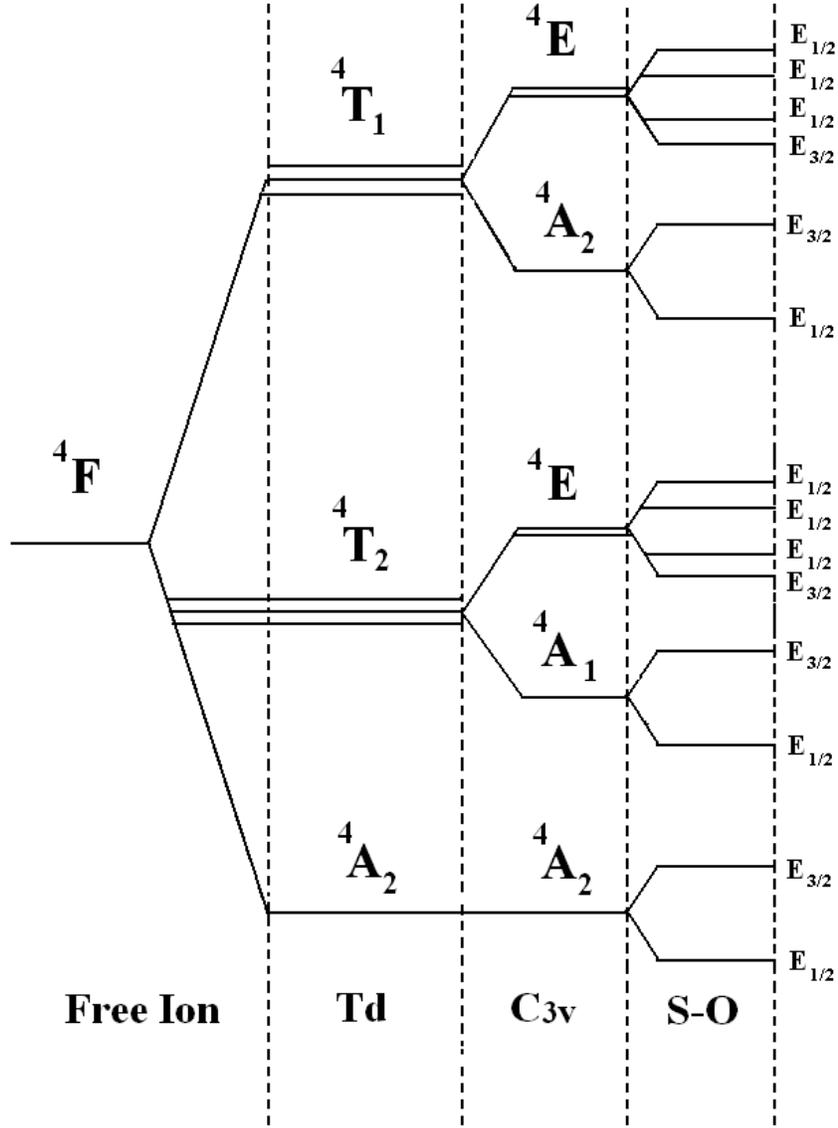}

      \caption{Qualitative splitting due to the cubic field,
      followed by the trigonal crystal field and by the SO effect.}\label{splitting_levels}
\end{center}
\end{figure}

For a while let us neglect the spin-orbit coupling. Then the total
spin $S$ and total angular momentum $L$ are conserved, because
Coulomb interaction is rotationally invariant and does not depend on
spin. The ground state for seven $d$-electrons (three holes) is
$\left| ^{4}F\right\rangle $ with $L=3,\ S=3/2$ and it is
$(2L+1)(2S+1)=28$-fold degenerate. The eigenfunctions are $\left|
L,M,m_s\right\rangle $, $M$ being the momentum projection on $z$
axis, $m_s$ the total spin projection. For $M=3$ and $m_s=\frac 32$
we have
\[
\left| 3,3,\frac 32\right\rangle =d_{0\uparrow }^{\dagger }d_{1\uparrow
}^{\dagger }d_{2\uparrow }^{\dagger }\left| vac\right\rangle .
\]
Acting on this state by $\hat{L}_{-}$ operator we obtain
successively
\[
\left| 3,2,\frac 32\right\rangle =d_{-1\uparrow }^{\dagger }d_{1\uparrow
}^{\dagger }d_{2\uparrow }^{\dagger }\left| vac\right\rangle ,
\]
\[
\left| 3,1,\frac 32\right\rangle =\left( \sqrt{\frac{2}{5}}d_{-2\uparrow
}^{\dagger
}d_{1\uparrow }^{\dagger }d_{2\uparrow }^{\dagger }+\sqrt{\frac{3}{5}}%
d_{-1\uparrow }^{\dagger }d_{0\uparrow }^{\dagger }d_{2\uparrow
}^{\dagger }\right) \left| vac\right\rangle ,
\]
\[
\left| 3,0,\frac 32\right\rangle =\sqrt{\frac{1}{5}}\left( 2d_{-2\uparrow
}^{\dagger }d_{0\uparrow }^{\dagger }d_{2\uparrow }^{\dagger
}+d_{-1\uparrow }^{\dagger }d_{0\uparrow }^{\dagger }d_{1\uparrow
}^{\dagger }\right) \left| vac\right\rangle ,\ etc.
\]
Under the action of cubic crystal field this level splits into 1
singlet and 2 triplets. Then the basis functions may be labeled as
$\left| 3,\chi ,\tilde{m},m_{s}\right\rangle $, where $\chi
=A_{2},T_{2},T_{1}$ denotes the representation of the cubic group,
$\tilde{m}$ is the projection of a fictive angular momentum within each
manifold. We have
\begin{eqnarray}
\nonumber
  \left| 3,A_{2} ,0,m_{s}\right\rangle  &=& -\frac{\sqrt{5}}{3}\left| 3,0,m_{s}\right\rangle +
  \frac{\sqrt{2}}{3}\left( \left|
  3,3,m_{s}\right\rangle -\left|
  3,-3,m_{s}\right\rangle \right)\\
\nonumber
  \left| 3,T_{2} ,\pm 1,m_{s}\right\rangle
  &=& \frac{1}{\sqrt{6}}\left| 3,\mp 2,m_{s}\right\rangle
  \pm
  \sqrt{\frac{5}{6}}\left| 3,\pm 1,m_{s}\right\rangle \\
\label{basis4F}
  \left| 3,T_{2} , 0,m_{s}\right\rangle  &=& \frac{1}{\sqrt{2}}\left( \left|
  3,3,m_{s}\right\rangle +\left|
  3,-3,m_{s}\right\rangle \right)  \\
\nonumber
  \left| 3,T_{1} ,\pm 1,m_{s}\right\rangle
  &=& \mp \sqrt{\frac{5}{6}}\left| 3,\mp 2,m_{s}\right\rangle
  +
  \frac{1}{\sqrt{6}}\left| 3,\pm 1,m_{s}\right\rangle  \\
\nonumber
  \left| 3,T_{1} , 0,m_{s}\right\rangle
  &=& -\frac{2}{3}\left| 3,0,m_{s}\right\rangle
  -
  \frac 13\sqrt{\frac{5}{2}}\left( \left|
  3,3,m_{s}\right\rangle -\left|
  3,-3,m_{s}\right\rangle \right)\ .
\end{eqnarray}
The trigonal field splits the triplets into doublets and singlets
and also couples the states of different manifolds with equal $
\tilde{m}$. This is schematically shown in Fig.\
\ref{splitting_levels}.
The lowest excited level is $^4P$ with $L=1,\ S=3/2$
\[
\left| 1,1,\frac 32 \right\rangle =\left( \sqrt{\frac
35}d_{-2\uparrow }^{\dagger
}d_{1\uparrow }^{\dagger }d_{2\uparrow }^{\dagger }-\sqrt{\frac 25}%
d_{-1\uparrow }^{\dagger }d_{0\uparrow }^{\dagger }d_{2\uparrow
}^{\dagger }\right) \left| vac\right\rangle ,
\]
\[
\left| 1,0,\frac 32 \right\rangle =\sqrt{\frac 15}\left(
d_{-2\uparrow }^{\dagger }d_{0\uparrow }^{\dagger }d_{2\uparrow
}^{\dagger }-2d_{-1\uparrow }^{\dagger }d_{0\uparrow }^{\dagger
}d_{1\uparrow }^{\dagger }\right) \left| vac\right\rangle ,\ etc.
\]
In the free ion it is separated by energy $15B$ from the ground
state. The cubic field couples the states $\left| 1,M,m_{s}\right\rangle
$ and $\left| 3,T_1,\tilde m,m_{s}\right\rangle $ with $\tilde m=M$.
The trigonal field has also matrix elements between
$\left| 3,A_{2},\tilde{m},m_{s}\right\rangle$,
$\left| 3,T_{2},\tilde{m},m_{s}\right\rangle $ and
$\left| 1,\tilde{m},m_{s}\right\rangle $ states that are proportional to
$v$ and $v^{\prime }$, thus they are much smaller.

\subsection{Perturbation theory}

 From the parameter values discussed below, it will become clear
that we are in the regime where $15B\gg \Delta \gg \upsilon ,\upsilon
^{\prime },\lambda $. In the strong cubic crystal field case (i.e.\
when $\Delta \gg 15B$) a perturbative formula for zero field
splitting $2D$ and the gyromagnetic factors $g_{\parallel }$ and
$g_{\perp }$ was developed by MacFarlane.\cite
{MacFarlane67,MacFarlane} But it is not adequate in the present
situation. Now, we can define four small values $\frac{\Delta
}{(15B)}$, $\frac{\upsilon }{\Delta } \sim \frac{\upsilon ^{\prime
}}{\Delta } \sim \frac{\lambda }{\Delta }$. In fact, we have $15B\gg
\Delta $, and the value $\Delta /15B $ being almost of the same order
of magnitude as $v/\Delta $, then the ratio
\[
\frac{v}{15B}=\frac{v}{\Delta }\frac{\Delta }{15B}
\]
can be considered as an order of magnitude smaller than $\Delta /15B $. This justifies
the application of the weak crystal field approach.

We will proceed in three steps that may be regarded as three
subsequent canonical transformations of our Hamiltonian similar to
Eq. (\ref{trans}). First, we eliminate the coupling with $^{4}P$
states retaining only the order $\Delta /15B$ (the explicit use of
weak CF scheme). Then only the states $\left|
3,T_{1},\tilde{m},m_{s}\right\rangle $ acquire an admixture
\begin{equation}
\left| T_{1},\tilde{m},m_{s}\right\rangle =\left| 3,T_{1},\tilde{m}%
,m_{s}\right\rangle +\frac{\left\langle 1,\tilde{m},m_{s}\right| \hat{H}%
_{CF}\left| 3,T_{1},\tilde{m},m_{s}\right\rangle }{\left(
E_{T_{1}0}-E_{P0}\right) }\left| 1,\tilde{m},m_{s}\right\rangle \ ,
\label{T1per}
\end{equation}
where
\[
\frac{\left\langle 1,\tilde{m},m_{s}\right| \hat{H}_{CF}\left| 3,T_{1},%
\tilde{m},m_{s}\right\rangle }{\left( E_{T_{1}0}-E_{P0}\right) }\simeq \frac{%
2}{5}\frac{\Delta }{15B}\equiv \kappa _{P}\ .
\]
We thus obtain an effective Hamiltonian acting in the $^{4}F$
subspace. In the next step we consider the perturbation due to the
trigonal field with the small parameters $\frac \upsilon \Delta$ and
$\frac{\upsilon^{\prime}}{ \Delta}$. In the following we will use
the first order ground state wave function of the approximately
diagonal crystal field Hamiltonian,\cite{exactCF} i.e.
\begin{equation}
\nonumber
  \left| \psi _{0}\right\rangle = \left| A_{2},0,m_{s}\right\rangle \approx
\left| 3,A_{2},0,m_{s}\right\rangle -\kappa _{0}\left|
T_{1},0,m_{s}\right\rangle \, ,
\end{equation}
as well as the excited states
\begin{eqnarray}
\nonumber
  \left| T_{2},0,m_{s}\right\rangle &\approx & \left|
3,T_{2},0,m_{s}\right\rangle \\
\label{psi1stord}
  \left| T_{2},\pm 1,m_{s}\right\rangle &\approx & \left| 3,T_{2},\pm
1,m_{s}\right\rangle \mp \kappa _{2}\left| T_{1},\pm
1,m_{s}\right\rangle \\
\nonumber
  \left| T_1,0,m_s\right\rangle _t &\approx & \left|
T_1,0,m_s\right\rangle
+\kappa _0\left| 3,A_2,0,m_s\right\rangle \\
\nonumber
  \left| T_1,\pm 1,m_s\right\rangle _t &\approx & \left| T_1,\pm
1,m_s\right\rangle
  \pm \kappa _2\left| 3,T_2,\pm
1,m_s\right\rangle \ ,
\end{eqnarray}
where
\begin{eqnarray}
\nonumber
\kappa _0\equiv \frac{\upsilon ^{\prime
}\sqrt{10}}{5\left( E_{T_10}-E_{A_2}\right) } \left( 1+2\kappa
_P\right) \simeq \frac{\upsilon ^{\prime }\sqrt{10}}{9\Delta }
\left( 1+2\kappa
_P\right) \ ,\\
\nonumber \kappa _2\equiv \frac{\upsilon +2\upsilon ^{\prime
}\sqrt{2}}{2\sqrt{5}\left( E_{T_11}-E_{T_21}\right) }+\frac{\kappa
_P}{\sqrt{5}\left( E_{T_11}-E_{T_21}\right) }\left(\upsilon
-\frac{3\sqrt{2} }2\upsilon ^{\prime }\right)\ .
\end{eqnarray}
The corresponding energies are
\begin{eqnarray}
\nonumber
  E_{A_2} &\approx & -\frac{6\Delta }{5} \\
\nonumber
  E_{T_20} &\approx & -\frac{\Delta }{5} +\frac{\upsilon }{3}\\
\label{E1stord}
  E_{T_21} &\approx & -\frac{\Delta }{5} -\frac{\upsilon }{6} \\
\nonumber
  E_{T_10} &\approx & \frac{3\Delta }{5} +\frac{3\upsilon }{5}+\frac{4\upsilon ^{\prime
}\sqrt{2}}{5} -\frac{4}{5}\frac{\Delta ^2}{75B}\\
\nonumber
  E_{T_11} &\approx & \frac{3\Delta }{5} -\frac{3\upsilon }{10}+\frac{\upsilon ^{\prime
}\sqrt{2}}{210} -\frac{4}{5}\frac{\Delta ^2}{75B}\ .
\end{eqnarray}

Then we consider the spin-orbit interaction as a
perturbation with respect to the crystal field Hamiltonian.
We thus obtain the usual formulae for the
$g$-factor and the anisotropy $D$
\begin{eqnarray}
g_{\mu \nu }-g_{s}=-2\lambda k\Lambda _{\mu \nu }\ ,  \label{gfactor} \\
D=-\lambda ^{2}\left(\Lambda _{zz}-\Lambda _{xx}\right)\ ,
\label{aniso}
\end{eqnarray}
where
\begin{equation}
\Lambda _{\mu \nu }=\sum_{n\neq 0}\frac{\left\langle \psi _{0}\right| \hat{L}%
^{\mu }\left| n\right\rangle \left\langle n\right| \hat{L}^{\nu
}\left| \psi _{0}\right\rangle }{E_{n}-E_{0}}\ ,  \label{pt}
\end{equation}
with $E_0 = E_{A_2} $ and $\lambda $ is defined in Eq. (\ref{SO4F}).
These are the parameters appearing in the effective spin
$(S=\frac{3}{2})$ Hamiltonian:
\begin{equation}\label{Hspin}
H_{spin}=\mu _{B}g_{\parallel }B_{z}S_{z}+\mu _{B}g_{\perp
}\left(B_{x}S_{x}+B_{y}S_{y}\right)+D\left[S_{z}^{2}-\frac 13
S\left(S+1\right)\right].
\end{equation}
Let us note that all energy denominators appearing in  $\Lambda
_{\mu \nu }$ (\ref{pt}) are of the order of cubic splitting $\Delta
$. Thus, the perturbation theory requires the SO coupling to fulfill $\lambda
\ll \Delta $. But $\lambda $ may be of the same order of magnitude
as $\upsilon$ and $\upsilon ^{\prime }$.
The operator $\hat L_z$ couples the ground state only with $\left|
T_2,0,m_s\right\rangle $
\[
\hat L_z\left| A_2,0,m_s\right\rangle =\left( 2+ \kappa _0\sqrt{5}
\right) \left| T_2,0,m_s\right\rangle \ ,
\]
then
\[
\Lambda _{zz}=\frac{4\left( 1+ \kappa _0\sqrt{5} \right) }{\left(
E_{T_20}-E_{A_2}\right) }\ .
\]
For $\hat L_x=(\hat L_{+}+\hat L_{-})/2$ we have
\[
\hat L_x\left| A_2,0,m_s\right\rangle =\frac 12\left\{
-2\sqrt{2}\left( \left| T_2,1,m_s\right\rangle -\left|
T_2,-1,m_s\right\rangle \right) \right.
\]
\[
- \kappa_0 \left. \left[ -\frac{3\sqrt{2}}2\left( \left|
T_1,1,m_s\right\rangle +\left| T_1,-1,m_s\right\rangle \right)
-\sqrt{\frac 52}\left( \left| T_2,1,m_s\right\rangle -\left|
T_2,-1,m_s\right\rangle \right) \right] \right\} \; ,
\]
and
\[
\Lambda _{xx}\simeq \frac{4\left( 1-\frac{\kappa _0\sqrt{5}}2\right) }{%
\left( E_{T_21}-E_{A_2}\right) }\ .
\]
The anisotropy constant appears only in the third order (second order of
spin-orbit and first order of trigonal field)
\[
D\approx \frac{4\lambda ^2}\Delta \left( \frac{\upsilon}{2 \Delta }-\frac{3\kappa _0%
\sqrt{5}}4\right) \approx \frac{\lambda ^2}{\Delta}\left[ 2 \frac{\upsilon}{\Delta} -\frac{10%
\sqrt{2}}3 \frac{v^{\prime }}{\Delta} \left( 1+2\kappa _P\right)
\right] .
\]
The final results are:
\begin{equation}
\begin{array}{ll}
D=\frac{\lambda ^{2}}{\Delta}\left[ 2 \frac{\upsilon}{\Delta} -\frac{10\sqrt{2}}{3}%
\frac{\upsilon^{\prime}}{\Delta}\left( 1+\frac{4}{75}\frac{\Delta
}{B}\right) \right] &
\\
g_{\parallel }=g_{s}-\frac{8\lambda }{\Delta } k \left[ 1-\frac{\upsilon }{%
3\Delta }+\frac{5\sqrt{2}}{9\Delta }\upsilon ^{\prime }\left( 1+\frac{4}{75}%
\frac{\Delta }{B}\right) \right] &  \\
g_{\perp }=g_{s}-\frac{8\lambda }{\Delta } k \left[ 1+\frac{\upsilon }{6\Delta }%
-\frac{5\sqrt{2}}{18\Delta }\upsilon ^{\prime }\left( 1+\frac{4}{75}\frac{%
\Delta }{B}\right) \right] \ .&
\end{array}
\label{pert}
\end{equation}
%where $\kappa=2\sqrt{10}\upsilon^{\prime}/9\Delta$.
%
An alternative perturbative formula for zero-field splitting $2D$,
which is valid in the present situation, was derived by Mao-Lu and
Min-Guang.\cite{mao} However, our result is much more compact than
theirs (the Eqs. (5)-(9) of Ref. \onlinecite{mao} take one page and
a half) and,  correspondingly, more practicable. We have checked
that the difference between the compact formulas (\ref{pert}) and
the result given in Ref. \onlinecite{mao} is very small and can be
neglected in numerical applications.

\begin{figure}[h]
    \begin{center}
      \includegraphics[scale=0.45,angle=-90]{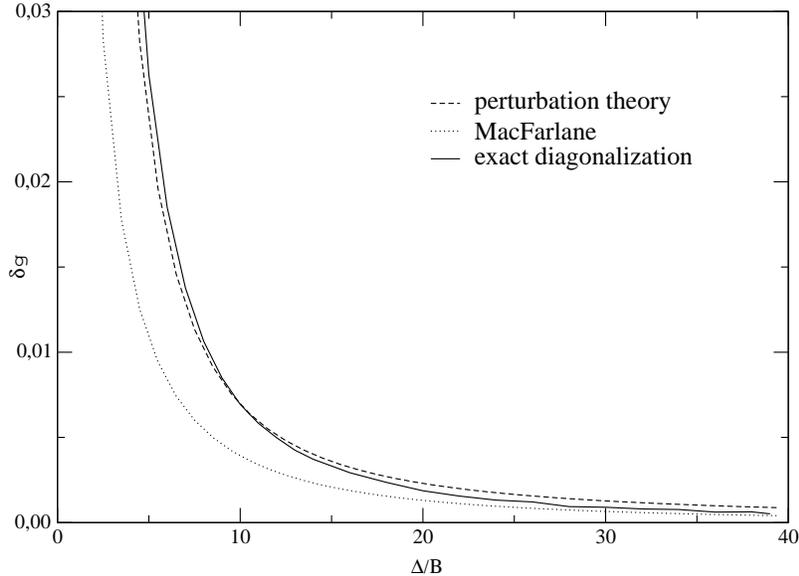}
      \caption{
Comparison between our perturbation theory (\ref{pert}) for the weak
field case (dashed line), the result derived by MacFarlane
\cite{MacFarlane} in the strong field case (dotted),and the exact
diagonalization calculation for $\delta g= g_{\perp} -
g_{\parallel}$ (full line). }\label{pertcomp}
\end{center}
\end{figure}

To check the applicability of our perturbation theory (\ref{pert})
we compared it with the exact diagonalization (see Appendix). We
fixed the parameters (in inverse centimeters) $\upsilon=54$,
$\upsilon^\prime=-213$, $C=3148$, $B=804$, $k=0.85$, and
$k\xi_{d,0}=481$ to those values which we derived for Co$^{2+}$ in
ZnO (see next Section) and varied $\Delta/B$ to display the
different regimes. As one can see from Fig.\ \ref{pertcomp}, the
difference of $g$-factors $\delta g = g_{\perp} - g_{\parallel}$
from our formula compares well with the exact diagonalization for
$\Delta/B$ up to 15. Therefore, we can certainly apply it to the
case of Co$^{2+}$ in ZnO where $\Delta/B$ is roughly 5 and where
there are considerable deviations for MacFarlanes formula.
\cite{MacFarlane} On the other hand, the exact diagonalization
converges towards MacFarlanes formula for large values of
$\Delta/B$.

\section{Application to ZnO:Co}

%\subsection{The parameters choice}

We apply our theory to the calculation of spin-Hamiltonian
parameters for Co impurity in zinc oxide. As input, we need the
following parameters of $\hat H$ (Eq. (\ref{Hpd})):
% \begin{itemize}
  (i) the structure of the Co environment given in Table
  \ref{Tab1},
  (ii) the charge transfer energy $\Delta _{pd}$ (Eq. (\ref{Deltapd2HF})),
  (iii) the free-ion spin-orbit coupling $\xi _{d,0}\approx  567\ {\rm cm}^{-1}$
  (see e.g. Table 7.6 of Ref. \onlinecite{abragam}),
  (iv) the Racah's parameters $B_0\approx 1115\ {\rm cm}^{-1}$ and $C_0\approx 4366\ {\rm cm}^{-1}$
(from Table 7.5 of Ref. \onlinecite{abragam}).
%\end{itemize}

As was mentioned above, we calculate the hopping integrals that
enter $\hat H$ (Eqs. (\ref{Hpd}) and (\ref{Tpd})) from Harrison's
expressions\cite{Harrison}
\begin{equation}\label{tpdm}
t_{pdm}(R)=\eta _{pdm}\frac{\hbar ^2r_d^{3/2}}{mR^{7/2}},\ \eta
_{pd\sigma }=-2.95,\ \eta _{pd\pi }=1.36,
\end{equation}
where the value $r_d=0.76\ \mathrm{\AA}$ for Co ion. The distance
$R$ is measured in {\AA} and $t_{pdm}$ in eV (1 eV = 8065.5
cm$^{-1}$). This gives, e.g. for $t_{pd\sigma }(R_4)\approx 1.34\
\mathrm{eV}$.
The coefficients $b_k$ (Eq. (\ref{bk})) are inversely proportional
to the charge transfer energy $\Delta _{pd}$. We choose the $\Delta
_{pd}\approx 3.6\ {\rm eV}\approx 28800\ {\rm cm}^{-1}$ so that the
cubic splitting $\Delta $ (Eq. (\ref{xyz2Dvvp})) is equal to the
experimentally determined value\cite{Koidl} $\Delta=4070$~cm$^{-1}$.
Then, the trigonal CF parameters $\upsilon \approx 54\ {\rm
cm}^{-1}$ and $\upsilon ^{\prime }\approx -213\ {\rm cm}^{-1}$ are
unambiguously determined by the Co environment via Eqs.
(\ref{xyz2Dvvp}) and {\em are not} additionally adjusted.
The SO and angular momentum reduction factor $k\approx 0.85$ was
{\em calculated} from Eq. (\ref{kred}). This gives the spin orbit
coupling $\xi _d =k\xi _{d,0}\approx 481\ {\rm cm}^{-1},\
\lambda=-\xi _d /3\approx -160\ {\rm cm}^{-1}$ very close to the
values met in the literature (and adjusted empirically): $\xi
_d=450\ {\rm cm}^{-1}$ in Ref. \onlinecite{MacFarlane}, and $\xi
_d=430\ {\rm cm}^{-1}$ in Ref. \onlinecite{Koidl}.
 From our above consideration, we have seen that the reduction
factors for Coulomb interaction may differ from those of SO and
angular momentum terms. Nevertheless, an order of magnitude estimate
may be done as $B\approx k^2 B_0\approx 804\ {\rm cm}^{-1}$ and
$C\approx k^2 C_0\approx 3148\ {\rm cm}^{-1}$. We adopt these values
for our calculations. The agreement with the experimentally adjusted
values $B=750$~cm$^{-1}$, $C=3500$~cm$^{-1}$ of
MacFarlane\cite{MacFarlane} is very good, keeping in mind the
roughness of the estimate. We should also note that the influence of
the $B$ and $C$ parameters on the calculated $g$-factors and the
zero field splitting $2D$ is rather small. So, using instead of our
estimations for $B$ and $C$ the values of MacFarlane (and keeping
all the other parameters constant) leads to $2D=4.882$,
$g_{\parallel}=2.230$, and $g_{\perp}=2.256$, quite close to the
results listed in Table II.

The parameters of the ground state spin-Hamiltonian (Eq.
(\ref{Hspin})) obtained by exact diagonalization of single-ion
Hamiltonian $\hat H_{eff}$ (Eq. (\ref{singleion})) are shown in the
Table \ref{Tab2}. Note that our sign convention for the crystal
field parameters corresponds to the electron representation in
contrast to the hole representation used in Refs. \onlinecite{Koidl}
and \onlinecite{MacFarlane}. The agreement with purely empirical
approaches, where all parameters are fitted to experiment, is very
good. We show also the parameters for the excited state.
\begin{table}
\begin{tabular}{|c|c|c|c|c|}
  \hline
    & P.\ Koidl\cite{Koidl} & R.M.\ MacFarlane\cite{MacFarlane} & Present work & experiment
    \\ \hline
     $\upsilon           $& -120 & -400 &   54 &  \\
     $\upsilon ^{\prime }$& -320 & -350 & -213 &  \\
                 $\Delta $& 4000 & 4000 & 4070 & 4070\cite{Koidl}  \\
                       $B$&  760 &  750 &  804 &  \\
                       $C$& 3500 & 3500 & 3148 &  \\
                       $k$&    1 &  0.8 & 0.85 &  \\
             $k\xi _{d,0}$&  430 &  450 & 481  &  \\ \hline
                      $2D$& 5.44 & 5.41 & 4.91 &  5.52\cite{Sati05} \\
          $g_{\parallel}$ & 2.24 & 2.20 & 2.23 & 2.236\cite{Sati05} \\
             $g_{\perp}$  & 2.28 & 2.23 & 2.26 & 2.277\cite{Sati05} \\ \hline
           $2D^{\prime }$ & 21.4 & 90.3 & 5.8  & 38\cite{Koidl} \\
$g_{\parallel}^{\prime }$ & 2.85 & 3.36 & 4.22 & 3.52\cite{Ferrand} \\
$E_{^2E\bar E}-E_{^4A\bar E}$&15171& 15050&14381&15123\cite{Ferrand}\\
  \hline
\end{tabular}
\caption{Measured ESR\cite{Sati05} and magneto-optic\cite{Ferrand}
data, compared to those calculated from CF theory with empirical
parameters\cite{Koidl,MacFarlane} or estimated from our approach.
$2D\ (2D')$ and $g\ (g')$ are the values for the ground (excited)
state. The energy unit is inverse centimeter. In the empirical CF
theory \cite{Koidl,MacFarlane} all the 7 parameters listed in the
upper part of the table are used as fitting parameters, whereas they
are microscopically calculated in our approach fixing only the cubic
CF splitting $\Delta$ to the experimental value. }\label{Tab2}
\end{table}

Table \ref{Tab3} compares the spin-Hamiltonian parameters obtained
by analytic perturbative approach (Eqs. (\ref{pert})) with the
results of the exact diagonalization and experiment. We see that the
analytic results lie within 15\% of accuracy. For the
phenomenological parameter set of Koidl\cite{Koidl} the accuracy is
about 20\%.\cite{Sati05} The reason is that in our set the absolute
values of trigonal parameters are smaller than for the set of Ref.
\onlinecite{Koidl} and the cubic splitting is the same, thus the
perturbation theory for our set converges better. The main reason of
the deviation from exact diagonalization is our neglect of the
interaction with $^2G$ term that violates the Hund's rule, but
nevertheless lies rather low in energy, just above the $^4P$ term
(see Table \ref{Tab4}). This deviation is remarkable for $2D$, but
much less for the $g$-factors.

\begin{table}
\begin{tabular}{|c|c|c|c|}
\hline
& experiment & diagonalization & perturbation theory \\ \hline
$2D$ & 5.52 & 4.91 & 4.31 \\
$g_{\parallel}$ & 2.24 & 2.23 & 2.25 \\
$g_{\perp}$ & 2.28 & 2.26 & 2.28 \\ \hline
\end{tabular}

\caption{ESR data as calculated from the numerical diagonalization and the
perturbation theory using the calculated crystal field parameters. \\[0pt]}\label{Tab3}
\end{table}

\section{Conclusion}

We have shown that Harrison's parametrization of electronic
structure of solids\cite{Harrison} may be successfully applied to
the calculation of spin-Hamiltonian parameters
(Eq. (\ref{Hspin})) for TMI
impurities in semiconductors (Eq. (\ref{pert})). It is especially useful
for the description of low symmetry paramagnetic centers as it
provides the connection between CF parameters and the geometry of
TMI surroundings (Eq. (\ref{xyz2Dvvp})). Thus, the number of empirically
adjustable parameters is substantially reduced.

We have demonstrated that the physical reason for the possibility to
apply the CF concept to TMI in semiconductors is the strong Coulomb
repulsion within the $d$-shell. It provides the large value of
charge transfer energy $\Delta _{pd}$
(Eqs. (\ref{Deltapd}) and (\ref{Deltapd2HF})) even in the case when the
mean-field energy of the $d$-level falls into the valence band. We
have given the explicit form of the canonical transformation
(Eqs. (\ref{trans}) and (\ref{W})) of the many-body Hamiltonian (Eq.(\ref{Hpd})) and
basis functions, which exploits this strongly correlated feature of
the TMI subsystem (Eq. (\ref{tllDelta})), and provides the effective
single-ion Hamiltonian (Eq. (\ref{Heff})). The latter connects the CF
Hamiltonian with the geometry of local surroundings of the impurity
and with the parameters of the electronic structure
(Eqs. (\ref{HCFgen}) to (\ref{bk})). The transformation
(Eqs. (\ref{trans}) and (\ref{W})) accounts for the covalency in the 'weak' CF
case within the Heitler-London configuration interaction approach. When applied to the
spin-orbit, Zeeman and Coulomb terms, it renormalizes their
parameters by covalency.

We have applied this theory to the Co impurity in ZnO. We have
adjusted only one parameter of our starting $p$-$d$ Hamiltonian
(Eq.\ (\ref{Hpd})), which acts in an energy scale of several eV. In
the result, we have fairly well reproduced a number of measurable
quantities available from ESR and optical experiments. Note that
these values reflect the tiny features of electronic structure
(magnetic anisotropy, Zeeman splitting), which have the scale of
several cm$^{-1}$. The results indicate that the proposed theory
catches the essential physics of TMI in semiconductors.

\section*{Acknowledgements}

The authors would like to thank A.~Stepanov, A.S.~Moskvin,
M.~D.~Kuzmin, M.~Richter, I.~Opahle, S.-L.~Drechsler, D.~Ferrand,
and J.~Cibert for many useful discussions. In part, this work was
supported by NATO Collaborative Grant CBP.NUKR.CLG 981255. R.O.K.
also acknowledge the support from the DFG project 436/UKR/17/8/05.

\section*{Appendix: Exact diagonalization for $3d^3$ ($3d^7$)}

In this section we detail the numerical calculations of the exact
diagonalization for 3 particles on a $3d$-level. The case of
interest namely $3d^7$ can be obtained from the former using a
particle-hole transformation.

The number of states for 3 particles on a $d$ level is
$10\times 9\times 8/3!=120$. These states can be labelled by usual quantum
numbers for total spin and angular momentum, we thus obtain
different multiplets: $^4F$  for which $S=\frac {3}{2}$, $L=3$, this
multiplet contains $(2S+1)(2L+1)=28$ different states. We also have
$^4P$ $(S=\frac {3}{2}, L=1)$, $^2H$ $(S=\frac {1}{2}, L=5)$, $^2G$
$(S=\frac {1}{2}, L=4)$, $^2F$ $(S=\frac {1}{2}, L=3)$, $^2D$ and
$^2D^{\prime }$ $(S=\frac {1}{2}, L=2)$, and finally $^2P$ $(S=\frac {1}{2},
L=1)$. This basis will be noted
$\{|\alpha \rangle=|S,L,M,m_s,\epsilon \rangle \}$, the last quantum number
$\epsilon$ is needed to distinguish states belonging to the 2
multiplets $^2D$ and $^2D^{\prime }$.

The $\{|\alpha \rangle\}$ basis is the natural one for the Coulomb
interaction as well as the Zeeman Hamiltonian, however an other
basis emerges when writing the one-body part of the Hamiltonian,
namely the crystal-field and spin-orbit terms. Let us denote by
$\{|i\rangle, i=1..10\}$, the basis for one electron on a $d$ level. $i$
is an index for the $(m_l,\sigma)$ state, where $m_l$ and $\sigma$
are the momentum and spin quantum numbers. We can construct a new
basis of 120 states $\{ |n\rangle \  = \ |ijk\rangle \ = \ c^{\dag}_i c^{\dag}_j
c^{\dag}_k |\ vac\rangle ,\ \  i<j<k: 1..10,\ \
 n: 1..120 \}$
where $|\ vac\rangle$ is the empty $d$ level.

The complete Hamiltonian has been written in Eq.
(\ref{singleion}). The Coulomb part is diagonal in the
$\{|\alpha \rangle\}$ basis, and the Tab. \ref{Tab4} gives the different
energy values.
\begin{table}
%\begin{center}
\begin{tabular}{|c|c|}
\hline  Term & Coulomb energy \\ \hline
$^4F$ & $3A -15 B$ \\
$^4P$ & $3A$   \\
$^2G$ & $3A -11 B +3C$\\
$^2H$ & $3A -6B +3C$\\
$^2P$ & $3A -6B +3C$ \\
$^2F$ & $3A +9 B  +3C$\\
$^2D,\ ^2D^{\prime }$ & $3A +5 B +5C \pm \sqrt(193B^2+8BC+4C^2)$ \\
\hline
\end{tabular}
\caption{ Coulomb energy depending on the multiplet}\label{Tab4}
\end{table}
Contrary to MacFarlane's work\cite{MacFarlane} where the Zeeman
Hamiltonian is treated in a perturbative manner, it is here
diagonalized on the same foot as the other terms. In the $\{|\alpha
\rangle\}$ basis, the Zeeman Hamiltonian is block-diagonal, and the
non zero matrix elements just connect states by $\hat{L}_{+(-)}$ or
$\hat{S}_{+(-)}$.
The crystal field Hamiltonian  is the sum of the cubic and trigonal
parts.
The one particle matrix elements of $\hat{H}_{CF}$ (Eq. \ref{HCFgen})) may
also be expressed in terms of Stevens equivalent operators
\cite{abragam}
\begin{equation}  \label{d1Stev}
\hat{H}_{cub}=-\frac{2}{ 3}B^{0}_{4}(\hat{O}_{4}^{0}-20\sqrt{2}
\hat{O}_{4}^{3}) \quad \hat{H}_{trig}= B^{^{\prime}}_{2} \hat{O}
_{2}^{0}+B_{4}^{^{\prime}}\hat{O}_{4}^{0}
\end{equation}
corresponding to a $d^1$ configuration.
The Stevens
operators are given by
\begin{eqnarray}
\nonumber
  \hat{O}_{4}^{0} & = &  35 \hat{L}_{z}^{4} -30 L(L+1) \hat{L}_{z}^{2} +25 \hat{L}_{z}^{2}
-6L(L+1)+3L^2(L+1)^2\\
\label{steve}
  \hat{O}_{4}^{3} & = & \frac {1}{4} \{ \hat{L}_{z} (\hat{L}_{+}^{3} +\hat{L}_{-}^{3} )
+  (\hat{L}_{+}^{3} +\hat{L}_{-}^{3} )\hat{L}_{z} \}\\
\nonumber
  \hat{O}_{2}^{0} & = &  3 \hat{L}_{z}^{2} - L(L+1) \ ,
\end{eqnarray}
where the operators $\hat{L}_{z}$,$\hat{L}_{+}$, or $\hat{L}_{-}$ are
\emph{one particle} operators.

In analytic calculations we have used the $\{|\alpha \rangle\}$
basis and for the $d^{7}$ configuration the crystal field
Hamiltonians (Eq.\ (\ref{d1Stev})) have to be used with the
parameters
\begin{equation}  \label{eq6}
\tilde{B}^{0}_{4}=-\frac{B^{0}_{4}}{5},\
\tilde{B}^{^{\prime}}_{4}=-\frac{B^{^{\prime}}_{4}}{5},\
\tilde{B}^{^{\prime}}_{2}=\frac{B^{^{\prime}}_{2}}{5}.
\end{equation}
The parameters $\Delta$, $\upsilon$, and $\upsilon^{\prime}$,
previously defined in Eq.\ (\ref{eq31}), are connected with the
Stevens parameters by:
\begin{equation}\label{Dvvp2Bd7}
\begin{array}{ll}
B^{0}_{4}=-\frac{\Delta}{120}-\frac{1}{360}(\upsilon+\frac{3
\sqrt{2}}{2}\upsilon^{\prime}) &  \\
B^{\prime}_{4}=-\frac{1}{140}(\upsilon+\frac{3\sqrt{2}}{2}
\upsilon^{\prime}) &  \\
B^{\prime}_{2}=\frac{\upsilon-2\sqrt{2}\upsilon^{\prime}}{21}
\; . &
\end{array}
\end{equation}
Using Eq. (\ref{xyz2Dvvp}) we obtain
\begin{eqnarray}
 \nonumber %to remove numbering (before each equation)
\tilde{B}_{4}^{0}
 &=&-\frac{\sqrt{6}}{2880}b_{4}(R_{1})\frac{z_1a^3}{R_1^4} \\
\label{bk2B} \tilde{B}_{4}^{\prime } &=& -\frac{1}{140}\left\{
b_{4}(R_{1})\frac{7\sqrt{6} z_1a^3+72(z_1^{4}-z_1^2a^{2})+3a^4}{216
R_1^4}+\frac{b_{4}(R_{4})}{9}
\right\}  \\
 \nonumber %to remove numbering (before each equation)
\tilde{B}_{2}^{\prime } &=& -\frac{1}{210}\left\{
b_{2}(R_{1})\frac{6z_1^{2}-a^2}{R_1^{2}} +2b_{2}(R_{4})\right\} ,
\end{eqnarray}
where $z_i$ is the $z$ coordinate of the ligand O$_i$;
$R_{1}=\sqrt{a^2/3+z_1^2}, \  R_4 = z_4$ are the corresponding
distances. For completeness we give here also the relation with
another parameter set, that is often met in the literature
\begin{equation}\label{Chin}
Dq=12B^0_4,\ D\tau =12B^{\prime }_4,\ D\sigma =-3B^{\prime }_2.
\end{equation}
This set is used e.g. in Ref. \onlinecite{mao}.

 The crystal field Hamiltonian is easily written
in the one-particle basis $\{|i\rangle, i=1..10 \}$, where one evaluates
the matrix elements $H^{CF}_{ij} = \langle i| H_{CF} | j\rangle$. Then the matrix
elements of the CF Hamiltonian can be written in the 3-particle basis $\{
|n\rangle = |ijk\rangle\}$ as follows
\begin{eqnarray}
\nonumber \langle k' j' i' | H_{CF} | i j k\rangle = H^{CF}_{i'i}
(\delta_{j'j} \delta_{k'k} -\delta_{j'k} \delta_{k'j}) -H^{CF}_{i'j}
(\delta_{j'i} \delta_{k'k} -\delta_{j'k} \delta_{k'i})
+H^{CF}_{i'k} (\delta_{j'i} \delta_{k'j} -\delta_{j'j} \delta_{k'i}) \\
\nonumber -H^{CF}_{j'i} (\delta_{i'j} \delta_{k'k} -\delta_{i'k}
\delta_{k'j}) +H^{CF}_{j'j} (\delta_{i'i} \delta_{k'k} -\delta_{i'k}
\delta_{k'i})
-H^{CF}_{j'k} (\delta_{i'i} \delta_{k'j} -\delta_{i'j} \delta_{k'i}) \\
\nonumber +H^{CF}_{k'i} (\delta_{i'j} \delta_{j'k} -\delta_{i'k}
\delta_{j'j}) -H^{CF}_{k'j} (\delta_{i'i} \delta_{j'k} -\delta_{i'k}
\delta_{j'i}) +H^{CF}_{k'k} (\delta_{i'i} \delta_{j'j} -\delta_{i'j}
\delta_{j'i})
\end{eqnarray}
where $\delta$ is the Kronecker symbol.
The spin-orbit term (Eq. (\ref{SO1ion})) is also easily written in
the one-particle basis $\{|i\rangle\}$, then in the 3-particle one $\{
|ijk\rangle\}$, using the preceding expansion.

At this stage we have some part of the total Hamiltonian written in
the $\{|\alpha\rangle\}$ basis, the other one in the $\{|ijk\rangle\}$ one. To
perform the numerical diagonalization, the last quantity needed is
the transformation matrix to connect these two basis. The
transformation basis is a kind of Clebsh-Gordan coefficient matrix
for three particles constrained by Pauli principle.


\begin{thebibliography}{99}
\bibitem{intermed} According another classification,\cite{Bethe,abragam} the 'weak' field for
3$d$ ions is called 'intermediate' field, and 'weak' field case
concerns $f$-ions, where cubic splitting is less than spin-orbit
coupling.
\bibitem{exactCF} In the $^4F$ subspace, the CF part of the Hamiltonian may be exactly solved analytically.
\cite{Villeret90} Unfortunately, this is not the case for  the
spin-orbit interaction matrix. As the SO coupling strength in the
$^{4}F$ subspace  $\lambda =-\xi _d/3 \sim \upsilon ,\upsilon \prime
$, we restrict our consideration to the lowest order of perturbation
theory.
\end{thebibliography}
\end{document}